\documentclass{PoS}
\usepackage[square,comma,numbers,sort&compress]{natbib}

\newcommand*{\La}{{\cal{L}}}
\newcommand*{\no}{\noindent}
\newcommand*{\bea}{\begin{eqnarray}}
\newcommand*{\eea}{\end{eqnarray}}
\newcommand*{\be}{\begin{equation}}
\newcommand*{\ee}{\end{equation}}
\newcommand*{\pd}{\partial}
\newcommand*{\pdm}{\pd_{\mu}}
\newcommand*{\pdn}{\pd_{\nu}}

\newcommand*{\pref}[1]{(\ref{#1})}

\newcommand*{\mn}{{\mu\nu}}

\newcommand*{\nn}{\nonumber}

\title{On the structure of the residual gauge orbit}

\ShortTitle{On the structure of the residual gauge orbit}

\author{\speaker{Axel Maas}\thanks{Supported by the DFG under grant number MA 3935/5-1.}\\
        Institute of Theoretical Physics, Friedrich-Schiller-University Jena, Max-Wien-Platz 1, D-07743 Jena, Germany\\
        E-mail: \email{axelmaas@web.de}}

\abstract{Gauge-fixed correlation functions are a valuable tool in intermediate steps when determining gauge-invariant physics. However, when obtaining them in different calculations, it is necessary to use exactly the same definition of the gauge to ensure comparability. Beyond perturbation theory, this is complicated by the Gribov-Singer ambiguity. In principle, lattice gauge theory can manipulate individual Gribov copies, thus making it an excellent method to deal with the ambiguity. However, to compare to continuum methods this requires to replicate the same treatment outside the lattice, usually in a path integral formulation. Here, the properties of the gauge orbit will be investigated with respect to this question. Especially, the possibility of employing averages over Gribov copies in non-perturbative generalizations of the Landau gauge will be discussed.}

\FullConference{International Workshop on QCD Green's Functions, Confinement and Phenomenology,\\
		September 05-09, 2011\\
		Trento Italy}

\begin{document}

\section{Non-perturbative gauge-fixing and the path integral}

In gauge theories like the standard model, gauge-fixed correlation functions, like the gluon propagator, are an excellent tool in intermediate steps to determine gauge-invariant physics. This has been heavily used both in perturbation theory \cite{Bohm:2001yx} and beyond \cite{Alkofer:2000wg,Fischer:2006ub,Binosi:2009qm,Maas:2011se,Boucaud:2011ug}. The key element in the latter calculations has been the judicious combination of lattice gauge theory, functional continuum methods, in particular Dyson-Schwinger equations (DSEs) and functional renormalization group equations (FRGs), effective theories, and perturbation theory \cite{Maas:2011se}. However, this requires to fix a gauge in a controlled way to determine these correlation functions. How to perform this beyond perturbation theory will be the central question in this work. For this purpose, it will be restricted to su(2) Yang-Mills theory in $d$ Euclidean space-time dimensions in the following.

In perturbation theory, gauge-fixing can be done in a unique way by employing local conditions, e.\ g.\ the Landau gauge condition used here
\be
\pdm A_\mu^a=0\label{lg}
\ee
\no on the gauge field $A_\mu^a$. It can be implemented in the path integral for the calculation of a quantity ${\cal O}$ using the Faddeev-Popov procedure \cite{Bohm:2001yx}
\bea
<{\cal O}>&=&\lim_{\xi\to 0}\int{\cal D}A_\mu{\cal D}c{\cal D}\bar{c} {\cal O}(A_\mu,c,\bar{c})e^{-\int d^4x {\La}_g}\label{pert}\\
{\La}_g&=&-\frac{1}{4}F_\mu^a F_{\mu a}+\frac{1}{2\xi}(\pdm A_\mu^a)^2+\bar{c}_a\pdm D_\mu^{ab} c_b\nn\\
F_\mn^a&=&\pdm A_\nu^a-\pdn A_\mu^a+g f^a_{bc} A_\mu^b A_\nu^c\nn\\
D_\mu^{ab}&=&\delta^{ab}\pdm+g f^{ab}_c A_\mu^c\nn
\eea
\no with the gauge-fixed Lagrangian $\La_g$ containing the coupling constant $g$, the structure constants $f^{abc}$, the (anti)ghost fields $c^a$ and $\bar{c}^a$, and the gauge parameter $\xi$, which has to be sent to zero at the end of the calculation.

The obstruction arising to this procedure beyond perturbation theory is the presence of multiple solutions to the original gauge condition \pref{lg}, the Gribov copies \cite{Gribov:1977wm}. The Gribov copies cannot be distinguished using any local condition in covariant gauges \cite{Singer:1978dk}. Such Gribov copies are just ordinary gauge copies, but now separated from each other by a finite gauge transformation. As such, they are also valid representatives of the gauge orbit, and it is required, just as with perturbative gauge copies, to identify a precise way how to deal with them. In particular, the same two possibilities are available as in perturbation theory \cite{Maas:2011se}: One possibility is akin to Landau gauge, and requires a (non-local) condition to select a well-defined representative for each gauge orbit among the Gribov copies. The alternative is to work akin to covariant gauges by averaging in a well-defined way over (a subset of) Gribov copies. As in perturbation theory, both possibilities are equally valid.

It remains how to perform such a task. In lattice gauge theory it is possible to determine at least a finite subset of the Gribov copies of each gauge orbit, and manipulate the Gribov copies directly. For the functional methods, this is not possible, and it is necessary to implement the prescription how to deal with the Gribov copies in another way. There appear to be at least two possibilities how to realize this \cite{Maas:2011se,Fischer:2008uz,Maas:2009se,Maas:2008ri,Maas:2010wb,Maas:unpublished}.

One method \cite{Maas:2011se,Fischer:2008uz,Maas:2008ri} starts by the realization that correlation functions determine a theory completely. Thus, two distinct gauges have to differ at least for one correlation function for one momentum configuration \cite{Maas:2009se}. Thus, without any restriction the equations for the correlation functions should contain all the different sets of correlation functions for the different gauges. Since the gauge-fixed Lagrangian itself generally changes along the gauge orbit, and thus for different Gribov copies, the necessary information is stored somewhere, and it should be possible to identify the different solutions by boundary conditions \cite{Fischer:2008uz}. Uniquely identifying suitable correlation functions and providing a formal proof of this statement is, however, complicated at best despite some tantalizing results \cite{Fischer:2008uz,Maas:2009se}. It thus remains an active area of research.

An alternative \cite{Maas:2011se,Maas:2010wb}, on which this work will be focused on, is to implement additional conditions in the path integral formulation similar to the perturbative Faddeev-Popov procedure. The by far simplest procedure can be formulated, if there exists a non-local quantity which is guaranteed to take values in a certain range along a gauge orbit for (almost all) configurations. Then, a replication of the perturbative procedure could be performed \cite{Maas:2010wb,Maas:2011se}. Whether such a quantity suited for Landau gauge exists is not proven. Thus, this option will not be investigated here. The alternative is instead to average over the gauge orbit with a well-defined weight. It is this possibility which will be explored here.

To simplify this, the first step is to reduce the path integral \pref{pert} to the so-called first Gribov region, i.\ e.\ the region where the Faddeev-Popov operator has only non-negative eigenvalues. It is known that this restriction does not affect any gauge-invariant quantity, as all gauge orbits have at least one gauge copy inside this region \cite{Dell'Antonio:1991xt}. The implementation of this in the path integral can be performed using a $\theta$-function as
\bea
<{\cal Q}>&=&\lim_{\xi\to 0}\int{\cal D}A_\mu{\cal D}c{\cal D}\bar{c} {\cal Q}\theta\left(-\pdm D_\mu^{ab}\right)(A_\mu,c,\bar{c})e^{-\int d^4x {\La}_g}\label{quant:fgr}\\
\theta\left(-\pdm D_\mu^{ab}\right)&=&\mathop{\Pi}_{i}\theta(\lambda_i)\nn,
\eea
\no where the $\lambda_i$ are the eigenvalues of the Faddeev-Popov operator $-\pdm D_\mu^{ab}$. Such an additional term, as stated, does not affect any gauge-invariant quantity. It can be implemented in lattice calculations \cite{Maas:2011se}. For functional calculations, the observation suffices that any variation of the such modified kernel of the path integral \pref{quant:fgr} will produce two terms. The first term will contain the variation of the $\theta$-function, yielding a $\delta$ function on the Gribov horizon. However, since on this horizon the Faddeev-Popov determinant, obtained after integrating out the ghost terms, vanishes, this term will not contribute \cite{Zwanziger:2003cf}. The second term produces only the ordinary functional equations, but with expectation values now obtained from the reduced integration domain of the first Gribov region. Turning the argument around, this implies that the functional equations over all Gribov regions have the same form, and thus have the same solution manifold. Thus, the solution manifold of these equations contain the correlation functions from both cases, and it requires a further constraint, a boundary condition, to select among them. Sufficient boundary conditions for this case are not yet identified, though some necessary ones are known \cite{Maas:2011se,Maas:2009se,Maas:2010wb}.

\section{The structure of the residual gauge orbit}

This, unfortunately, is not sufficient to fully eliminate Gribov copies \cite{vanBaal:1997gu}, and a possibly infinite number of them remains. This set is termed the residual gauge orbit \cite{Maas:2008ri}. In the following, lattice calculations will be used to investigate this residual gauge orbit further, before returning to a formulation in terms of the path integral. This implies the presence of both finite volume and discretization artifacts, and thus, strictly speaking, the following results only apply to the investigated set of lattice parameters.

In lattice calculations multiple restart algorithms \cite{Cucchieri:1997dx,Maas:2011se} permit to obtain at least a fraction of all Gribov copies of the residual gauge orbit. However, to distinguish two Gribov copies from each other in practical calculations is a non-trivial problem. In principle, it would be necessary to check at each space-time point for differences, within the numerical accuracy, taking all possible global transformations into account. Already the memory limitations are prohibitive in practice for any appreciable number of Gribov copies. Thus, an obtained gauge-fixed configuration is usually \cite{Maas:2011se} characterized by a finite number of moments. In actual calculations, these are based on the simplest ones, the two-point correlation functions,
\bea
F&=&1-\frac{1}{V}\int d^d x A_\mu^a A_\mu^a\nn\\
b&=&\tilde{Z}_3\lim_{p^2\to 0}G(p)\nn,
\eea
\no where $G$ is the color-averaged ghost dressing function and $\tilde{Z}_3$ is the ghost renormalization constant. Note that both quantities are evaluated for each Gribov copy and configuration individually. The finite volume in lattice calculations imposes that $b$ is actually evaluated at a finite momentum of order $1/L$ where $L$ is the lattice extension \cite{Maas:2011se}. This will here be considered as a finite volume artifact \cite{Maas:2009se,Maas:2009ph}. While $F$ as a composite operator is subject to well-defined additive and multiplicative renormalizations \cite{Itzykson:1980rh}, $b$ is only multiplicatively renormalized. Note that both quantities are invariant under global color rotations and space-time transformations.

In the following, two Gribov copies will be considered to be distinct if their difference in $F$ or $b$ exceeds a certain threshold $\epsilon$. Hence, some Gribov copies, which are different, will not be recognized as such. Thus, all results which indicate a sensitivity to the choice of Gribov copies yields only a lower limit to this sensitivity, even if all Gribov copies of any given residual gauge orbit would be determined. So, the value of $\epsilon$ has to be set carefully. For a properly renormalized quantity, $\epsilon$ could be the numerical accuracy, which is usually significantly less than the machine precision, giving possible cancellations in the sums and imprecision in the gauge-fixing process. For $b$, this is rather straight-forward. For $F$, this is less simple. Due to the combination of additive and multiplicative renormalization, the unrenormalized $F$ will tend towards 1 in the continuum limit with vanishing width. Thus, at fixed numerical accuracy two distinct Gribov copies at low discretization will appear equal at better discretization, because their difference in $F$ will no longer be resolvable.

\begin{figure}
\includegraphics[width=0.5\textwidth]{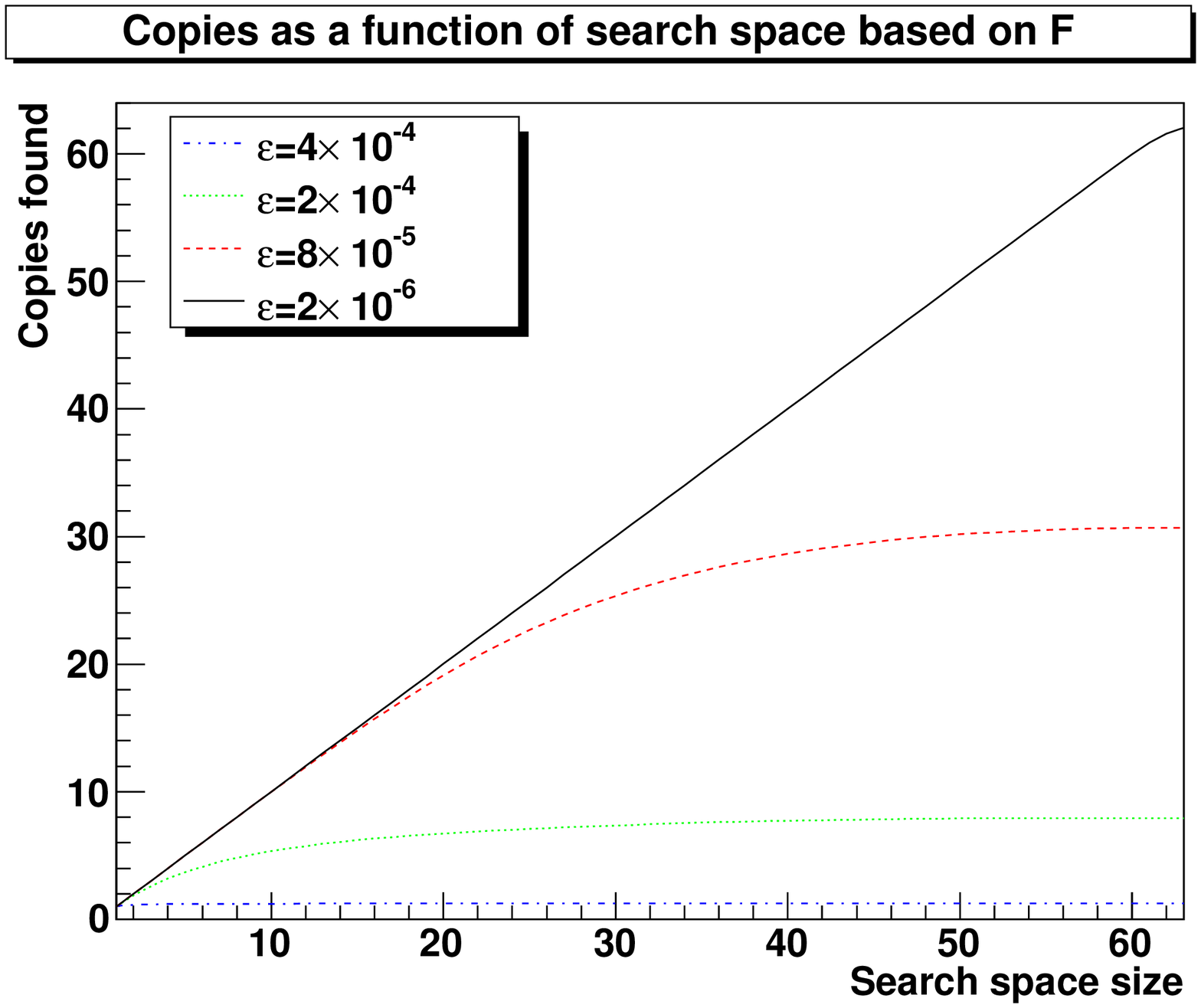}\includegraphics[width=0.5\textwidth]{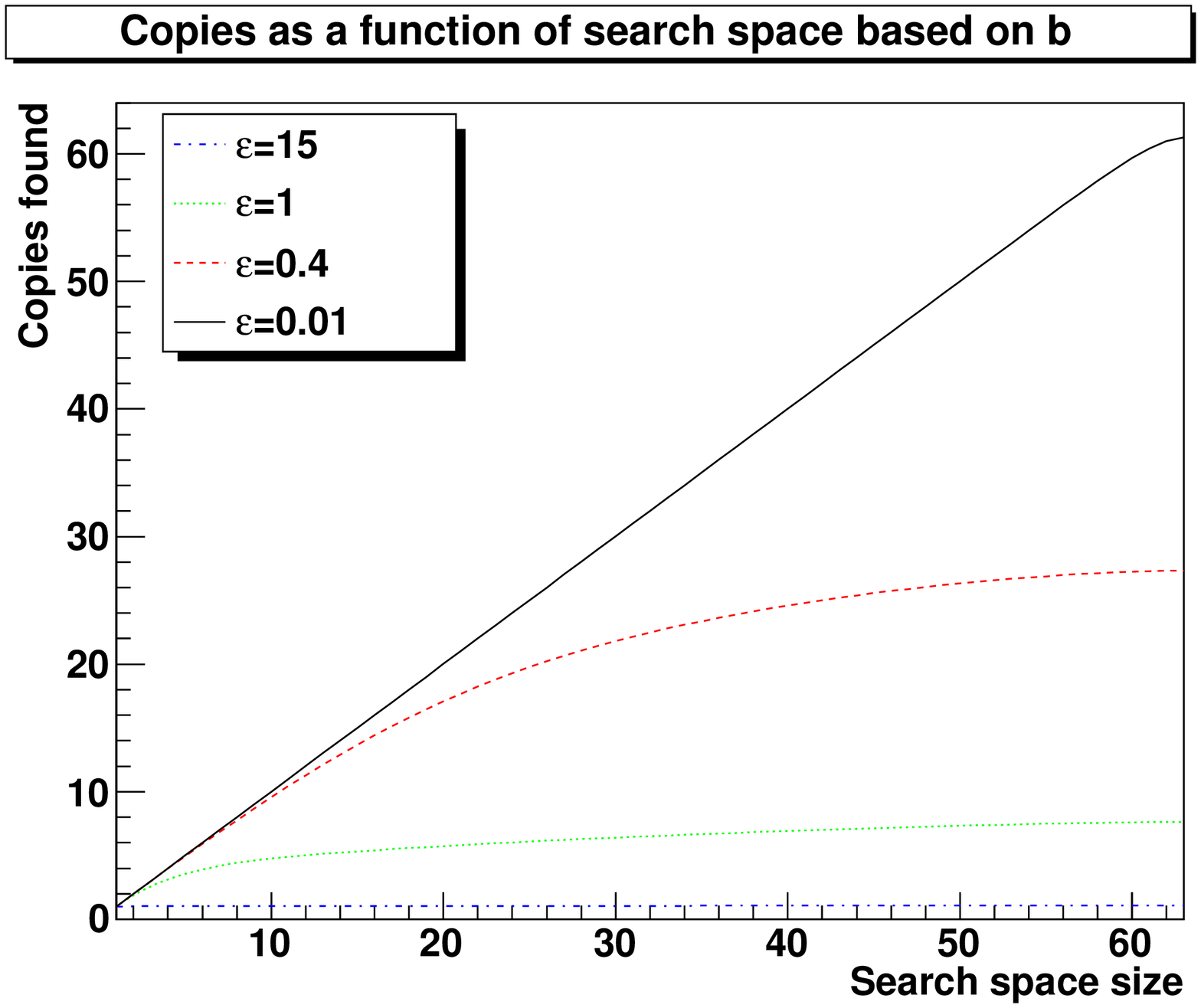}
\caption{\label{cpnr}The number of Gribov copies identified, both as a function of the search space size and the threshold $\epsilon$. Results are from a (4.4 fm)$^3$ lattice with lattice spacing 0.122 fm in three dimensions. Results in the left panel are obtained using $F$ to distinguish Gribov copies, and $b$ is used in the right panel.}
\end{figure}

To illustrate the impact of both restricting effects, the finite resolution of Gribov copies and the finite search space, the number of Gribov copies obtained using either $F$ or $b$ as the sole criterion is shown in figure \ref{cpnr}. From the figure it can be seen that the number found depends strongly on the threshold $\epsilon$ and the search space size, but not on whether either $F$ or $b$ is chosen.

Thus, both coordinates $F$ and $b$ vary between Gribov copies. At this point, an interesting question is whether Gribov copies resolved by either of the coordinates are also resolved by the other coordinate.  There are indications that this is the case \cite{Maas:2009se}. But this is not an entirely simple question, as this may depend on the resolving power with a fixed $\epsilon$. Thus this may be altered by a more systematic study, which is a rather non-trivial exercise, and will be done elsewhere \cite{Maas:unpublished}. However, for the present purpose, this is not relevant, as even this will not eliminate the possibility that two copies are not resolved by either $b$ or $F$, but only by some coordinate based on some higher order correlation function or a different kinematic definition.

\begin{figure}
\includegraphics[width=0.5\textwidth]{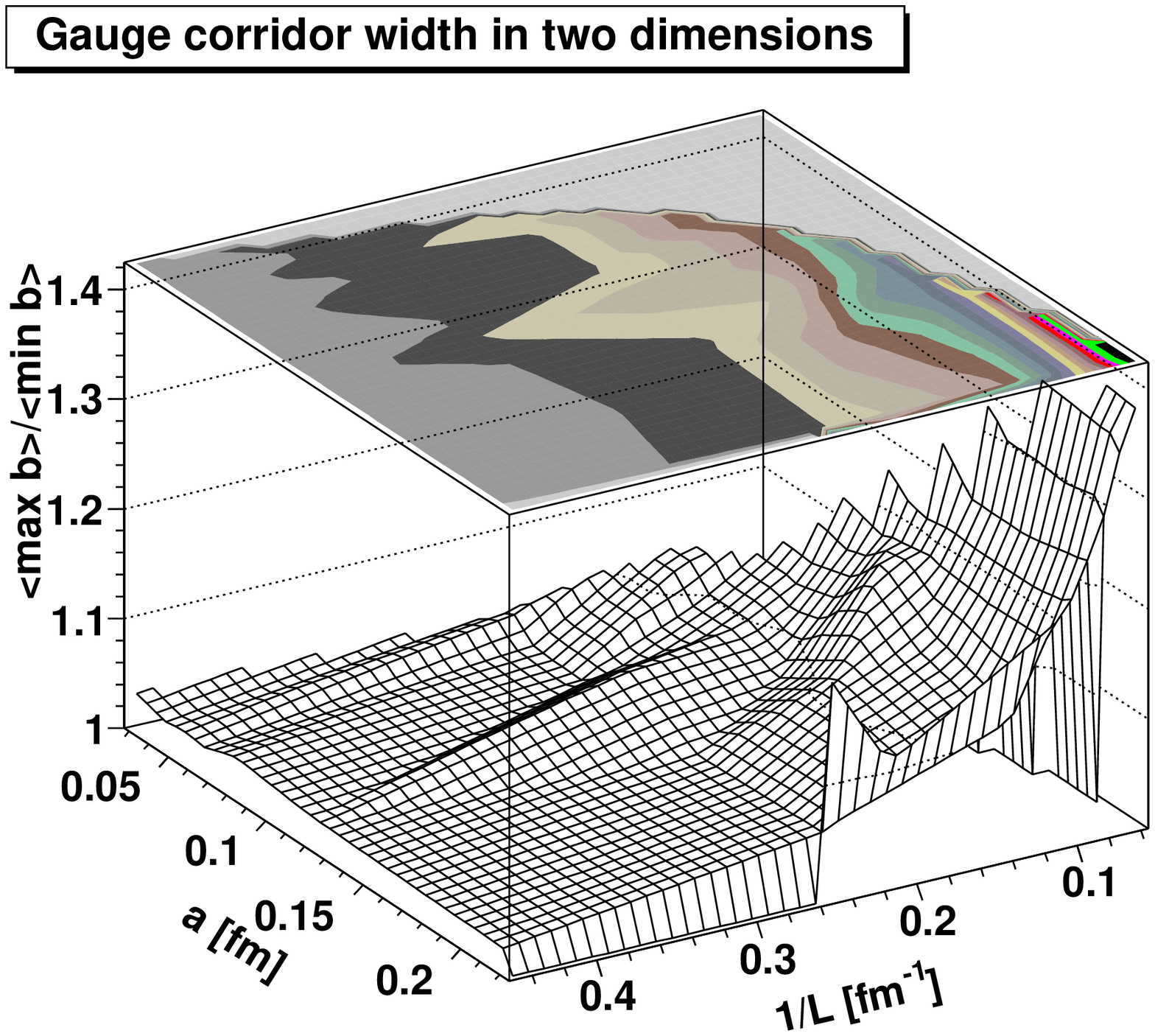}\includegraphics[width=0.5\textwidth]{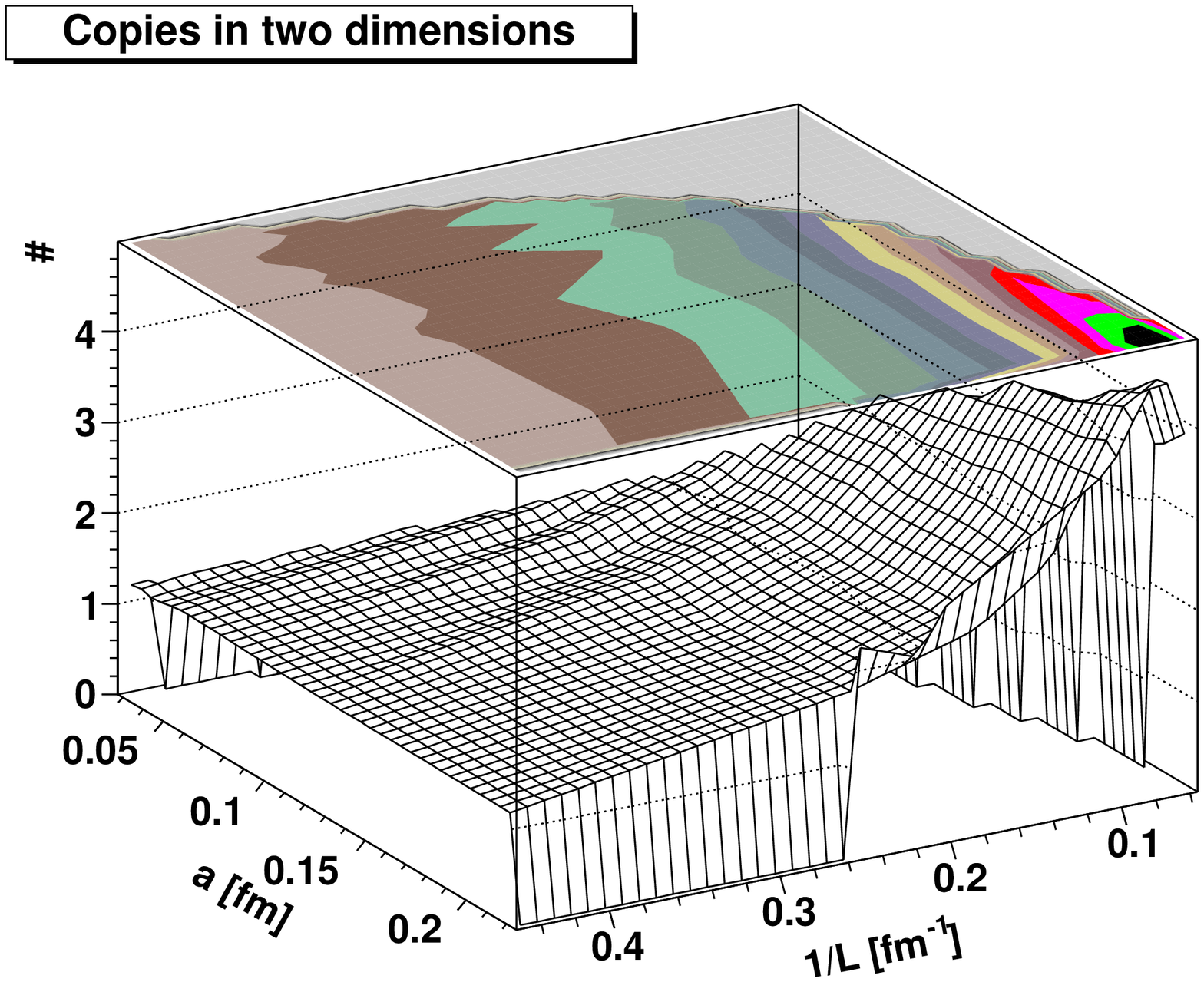}\\
\includegraphics[width=0.5\textwidth]{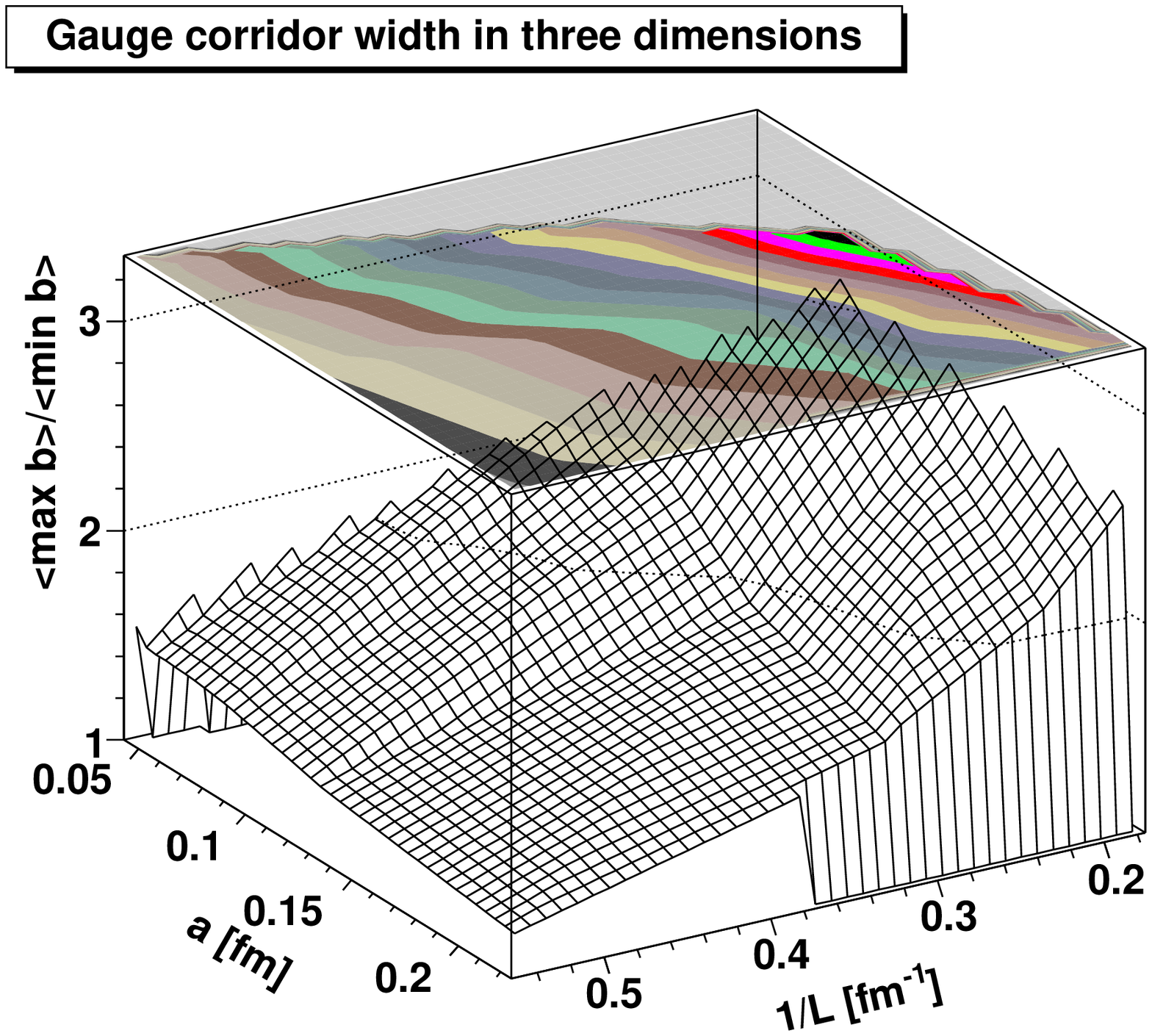}\includegraphics[width=0.5\textwidth]{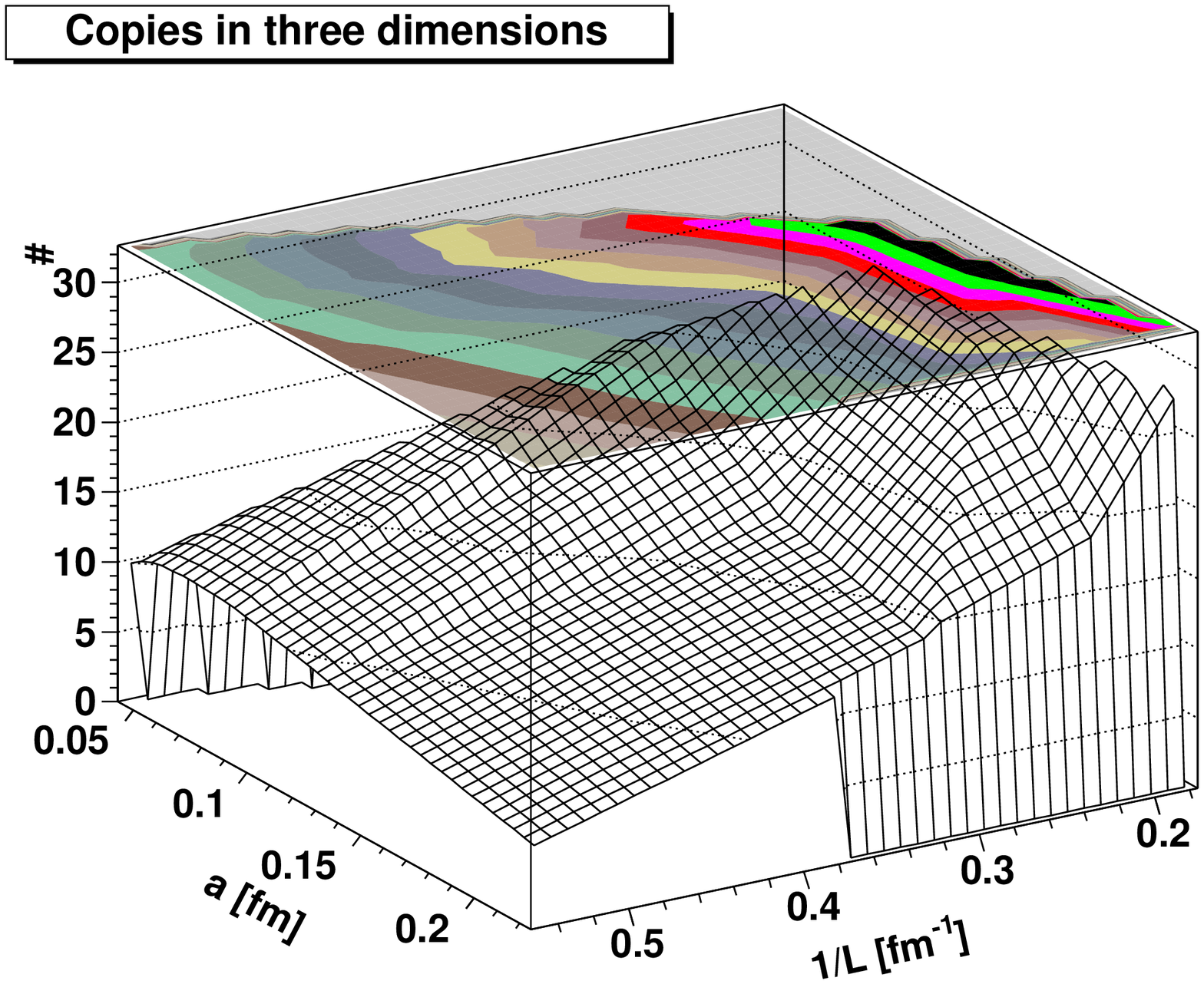}\\
\includegraphics[width=0.5\textwidth]{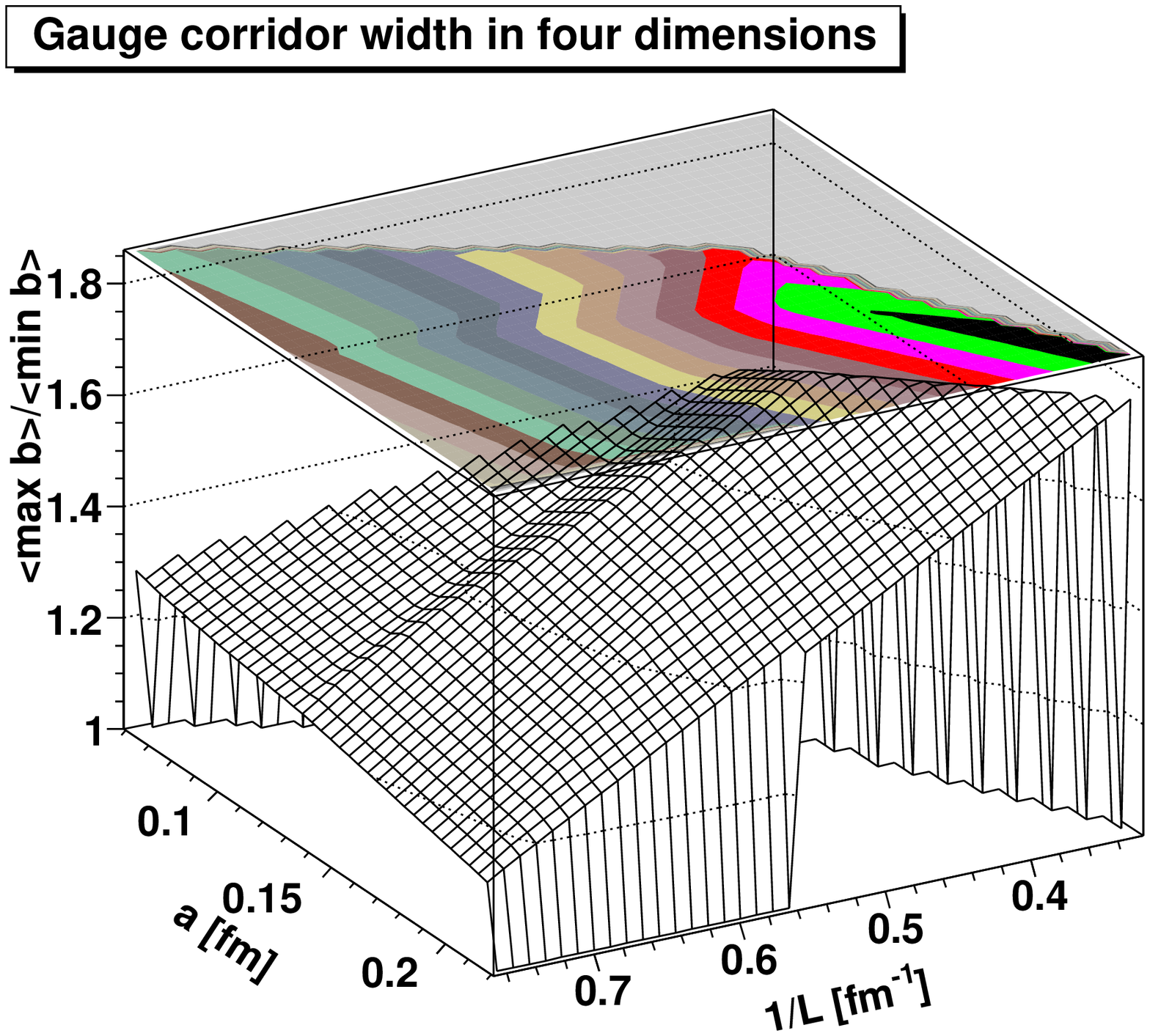}\includegraphics[width=0.5\textwidth]{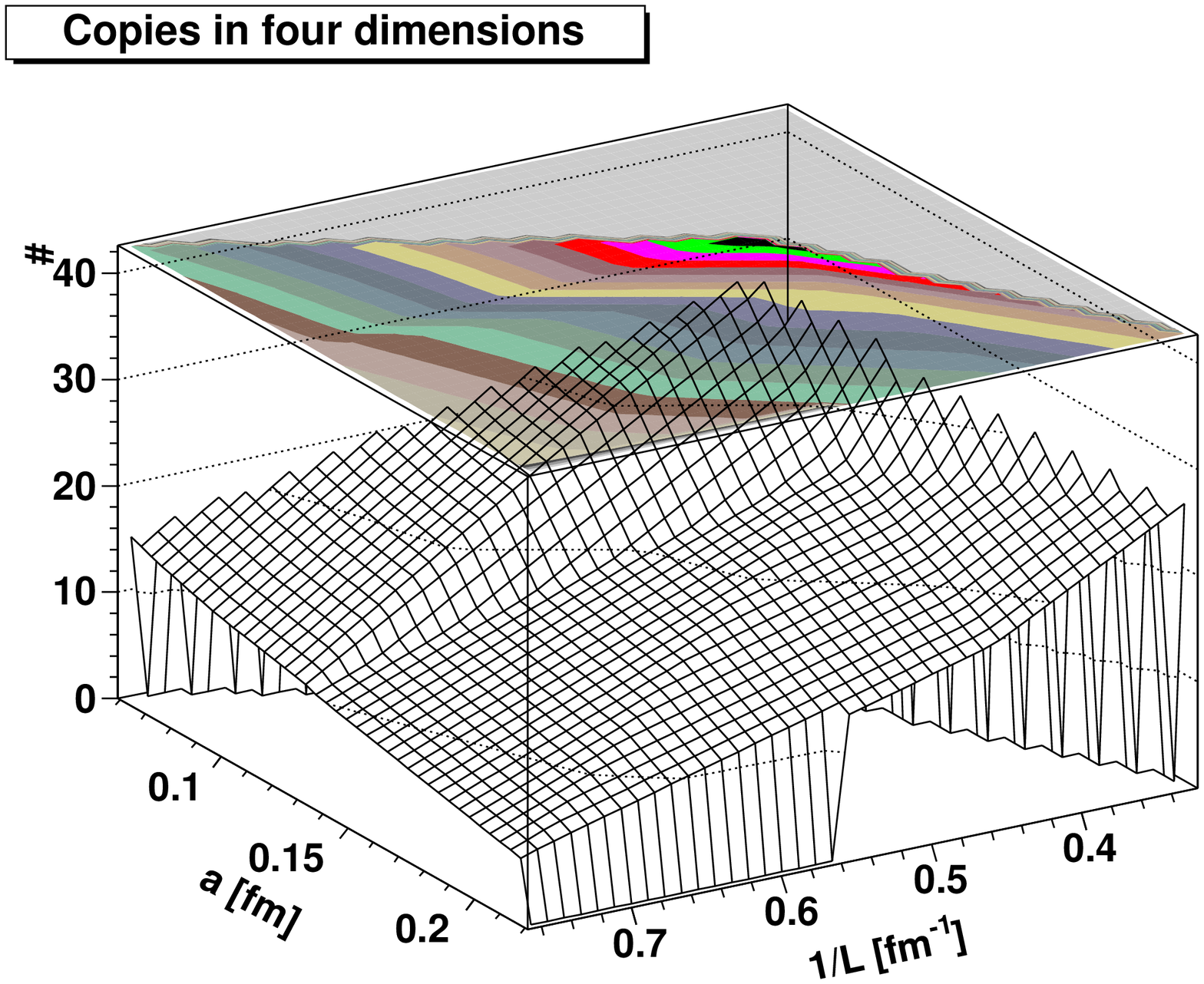}
\caption{\label{counts} Left panels: The ratio $\langle\max b\rangle/\langle\min b\rangle$ as a function of volume and lattice spacing in two (top panel), three (middle panel) and four (bottom panel) dimensions. Right panels: The number of Gribov copies as a function of volume and lattice spacing in two (top panel), three (middle panel) and four (bottom panel) dimensions. More details will be available in \cite{Maas:unpublished}.}
\end{figure}

In fact, the only relevant question in the following to obtain non-trivial results will be whether either of the coordinates has any resolving power at all, in particular towards the continuum and infinite-volume limit. This will be studied here for the coordinate $b$ by using two criteria. One is how the possible range of $b$ values develops and the other how the number of Gribov copies develops. The first quantity has an implication for the second quantity: Only if the range remains at least finite, it makes any sense to distinguish Gribov copies using $b$. This range, given as a ratio to remove all trivial multiplicative factors, is shown in the left-hand side of figure \ref{counts}. So far, it seems to be finite, and except for one single lattice setup in four dimensions not shrinking. Whether the latter is an artifact of too small a search space has to be seen, but it could also be a genuine effect. For now, it suffices that this range is finite, so it seems to be useful to differentiate between at least a part of the Gribov copies. Even if not, the whole procedure developed here could also be performed using a more suitable quantity.

To take renormalization effects into account, the parameter $\epsilon$ for counting the Gribov copies is then not fixed for all volumes and discretizations, but defined to be 1/1000th of the difference between $\langle\max b\rangle$ and $\langle\min b\rangle$. This implies that at most 1000 different Gribov copies can be found, and in particular close-by Gribov copies will not be resolved. Thus, the number of Gribov copies found is once more a lower limit. The results are shown in the right-hand side of figure \ref{counts}. In contrast to the original anticipation \cite{Maas:2008ri}, the Gribov copies show a different behavior in two dimensions than in three and four dimensions. While in the latter cases the number of Gribov copies increases as a function of $1/a$, in two dimensions this number is more or less constant. Furthermore, for the largest volume a decrease seems to occur in two dimensions, but since this is a result from a single lattice setup, this may or may not be an artifact. Nonetheless, the whole observation may stand in connection with the qualitative difference observed for the behavior of propagators in two and higher dimensions \cite{Cucchieri:2011ig,Maas:2011se}. 

\section{Gribov-copy-aware gauge-fixing using the path integral}

Assume for the following that the gauge corridor for $b$ is finite in the infinite-volume and continuum limit. The implications if this is not the case will be discussed afterwards.

This assumption implies that (at least some of) the Gribov copies are differentiated by their value of $b$. It is thus a viable concept to average over the Gribov copies with an additional weight factor, not only the Faddeev-Popov term, sensitive to the different values of $b$ for each Gribov copy. Formally, this can be justified by an argument that there exists an (unknown) functional which reduces the path integral exactly to one Gribov copy for each gauge orbit. Afterwards, an averaging over such reduced path integrals can be done, just as for covariant gauges in perturbation theory \cite{Bohm:2001yx}. The lattice implementation gives an operational definition of selecting a single Gribov copy, showing that this is possible. It is then permissible to average such expressions with an appropriate normalized weight, as long as this guarantees that any gauge-invariant operator remains unchanged. Otherwise, the weight is arbitrary. Such averages are, again just as for perturbative covariant gauges, merely a gauge choice.

A second constraint is that this does not alter the corresponding functional equations severely, if possible. This is not necessary, but technically convenient. A possibility may be then to write a weight factor explicitly depending on $b$ \cite{Maas:2011se,Maas:2010wb}
\bea
<{\cal O}>&=&\lim_{\xi\to 0}\int{\cal D}A_\mu{\cal D}c{\cal D}\bar{c} {\cal O}(A_\mu,c,\bar{c})\theta\left(-\pdm D_\mu^{ab}\right)e^{-\int d^4x {\La}_g}\nn\\
&&\times\exp\left({\cal N}+\lambda\frac{1}{V}\int d^dxd^dy\pdm^x\bar{c}^a(x)\pdm^yc^a(y)\right)\label{quant:bgauge2},
\eea
\no which only modifies the equation for the ghost propagator, since the additional term is bilinear in the fields. The parameter $\lambda$ is a Lagrange parameter, and can be seen as an additional gauge parameter. The quantity ${\cal N}$ is a normalization factor, which guarantees that this expression does not change gauge-invariant quantities. The double-integral is the ghost dressing function at zero momentum \cite{Maas:2011se}, and thus should act only as a boundary term in a functional equation.

It remains to investigate whether this expression could make sense. For this purpose, it can be operationally defined in lattice calculations, by a weighted sum over the Gribov copies of each residual gauge orbit with the exponential factor $exp(-\lambda b+{\cal N})$, with ${\cal N}$ chosen such that the expectation value of one remains one. Since this then only averages over gauge copies, this is nothing but a gauge choice in lattice calculations, and therefore admissible.

\begin{figure}
\includegraphics[width=0.5\textwidth]{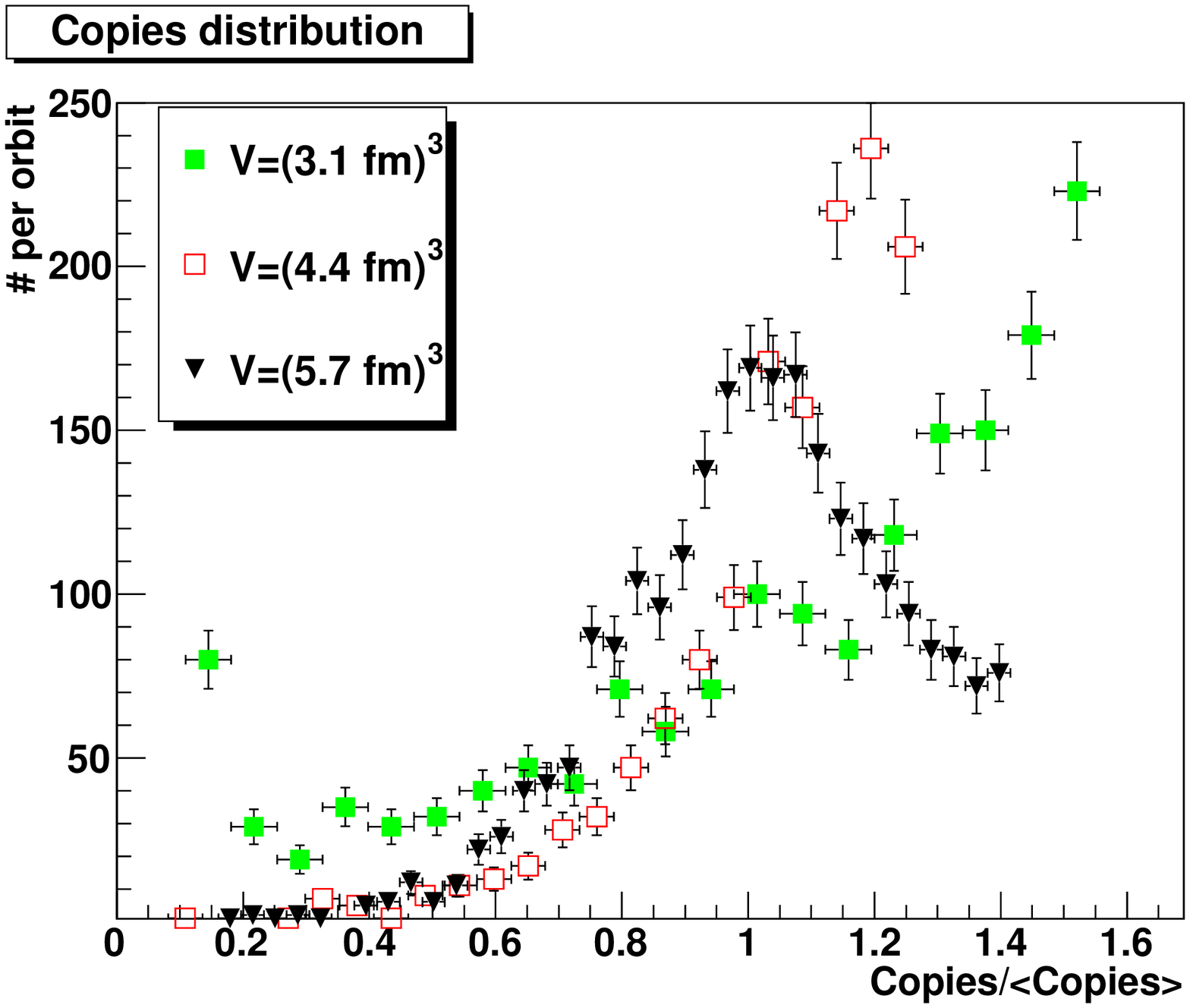}\includegraphics[width=0.5\textwidth]{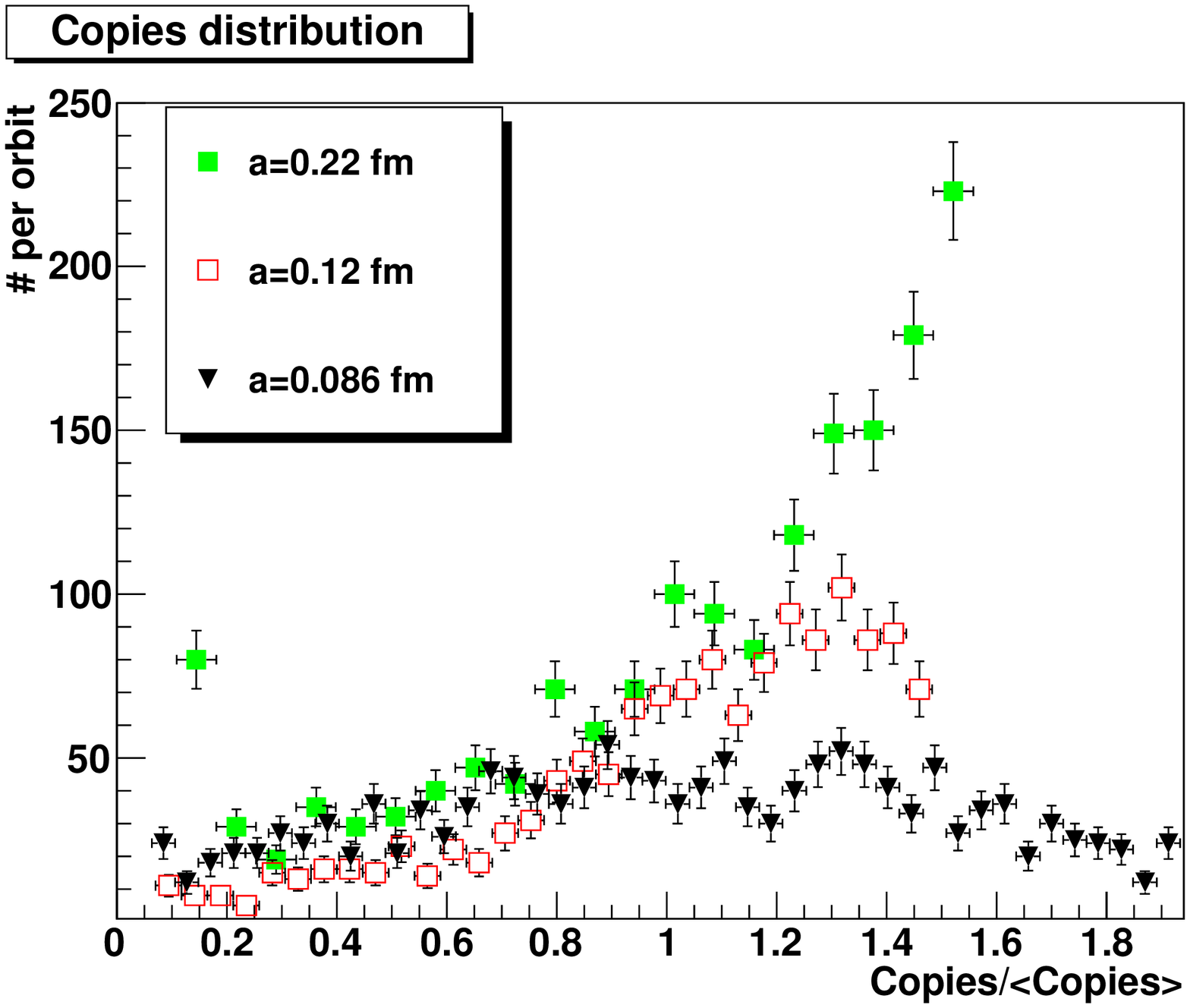}
\caption{\label{gdis}The distribution of Gribov copies over the orbits in three dimensions, normalized to the average number of Gribov copies. The left-hand side shows the distribution as a function of volume at fixed discretization $a=0.22$ fm, the right-hand side as a function of discretization at fixed $V=(3.1$ fm$)^3$ \cite{Maas:unpublished}.}
\end{figure}

The first interesting question is, whether the normalization constant ${\cal N}$ is gauge-orbit independent, at least in the infinite-volume and continuum limit. If this were the case, it could be absorbed in the measure. The simplest form for this will evidently be obtained for $\lambda=0$. In this case, ${\cal N}$ merely counts the Gribov copies. Though this is not sufficient for ${\cal N}$ to be orbit independent, it is surely necessary. To investigate this, its distribution for different volumes and discretizations is shown in figure \ref{gdis}. The result is rather puzzling. With increasing volume, the distribution becomes better centered around the average value, and less asymmetric. However, for better discretization, the distribution becomes at the same time broader, but also less asymmetric. The eventual limit cannot be inferred from these figures, but instead calls for a more detailed investigation to be presented elsewhere \cite{Maas:unpublished}. Especially, the restricted search space and Gribov copy separating power could have a systematic influence here as well. Sure is only that for the present calculations the quantity ${\cal N}$ cannot be assumed to be orbit-independent.

\begin{figure}
\includegraphics[width=\textwidth]{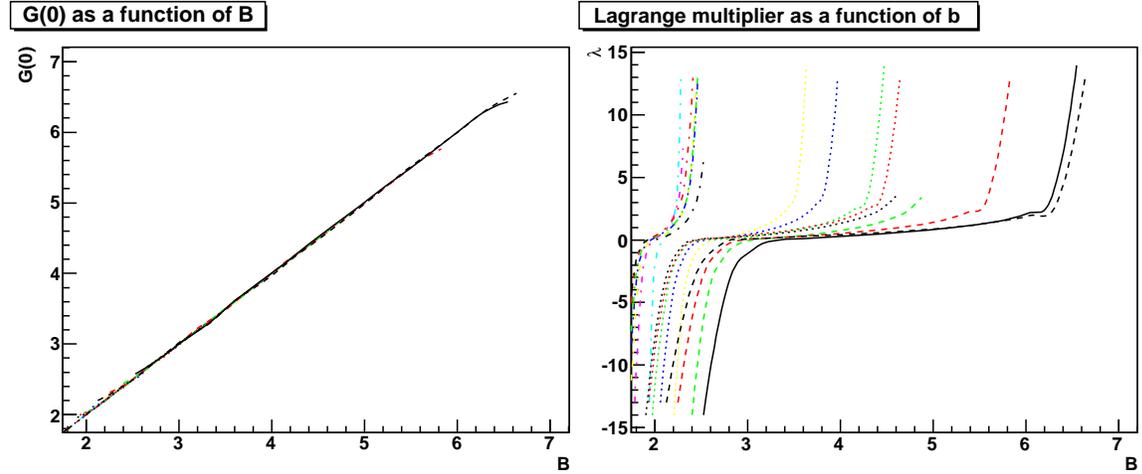}
\caption{\label{gvsb}The left panel shows the actual obtained value of $G(0)$ as a function of the desired $B$ values. The right-hand panel shows the value of $\lambda$ required to obtain the equality, up to numerical precision, shown in the left-hand side as a function of the gauge parameter $B$. The results are in three dimensions for volumes varying between (1.8 fm)$^3$ and (5.7 fm)$^3$ and for discretizations between $a=0.22$ fm to $a=0.043$ fm.}
\end{figure}

This is not a problem in a lattice calculation, and thus it is possible to proceed. To specify $\lambda$, there are two options. The first is to directly specify $\lambda$, as with the gauge parameter in covariant gauges. The alternative is an indirect condition, which will be done here along the lines of the Landau-$B$ gauge construction \cite{Maas:2009se}. In this case, the gauge condition $\langle b\rangle=B=G(0)$ for some value of $B$ is imposed, and a value of $\lambda$ is selected such that this condition is satisfied. This is indeed possible, as is shown in figure \ref{gvsb}. For a value of $B$ between $\langle\min b\rangle$ and $\langle\max b\rangle$ it is, within the numerical precision applied\footnote{It was searched for a fitting $\lambda$ value in the interval $\left[-14,14\right]$ with step size 0.2.}, always possible to find a value of $\lambda$ such that the prescription \pref{quant:bgauge2} yields the desired equality. Since the weighted average cannot exceed the values of $\langle\min b\rangle$ and $\langle\max b\rangle$, there is no $\lambda$ value which could permit to go beyond this range. As a consequence, when $B$ approaches either of the values, the required $\lambda$ value diverges, clearly marking the impossibility to exceed these bounds. On the other hand, this also shows that choosing a  particular value of $\lambda$ just leads to a particular value of $\langle b\rangle$, and both conditions are in a one-to-one correspondence.

\begin{figure}
\begin{center}\includegraphics[width=0.5\textwidth]{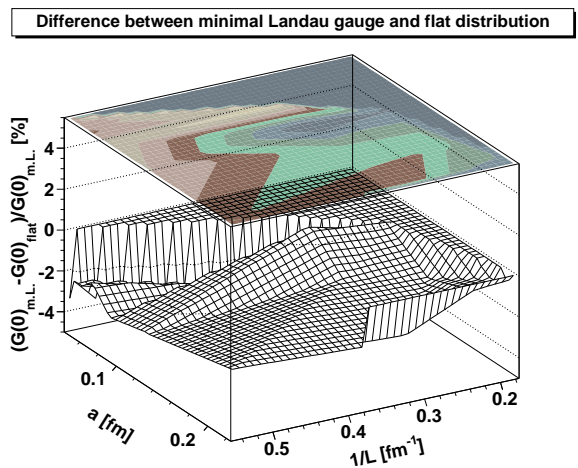}\end{center}
\caption{\label{diff}The difference in percent in $G(0)$ between the minimal Landau gauge and an averaging with $\lambda=0$ in three dimensions.}
\end{figure}

An interesting case is $\lambda=0$, which corresponds to a flat average over the residual gauge orbit. If a random selection of a Gribov copy, as is done in the minimal Landau gauge \cite{Maas:2011se}, is an unbiased process, this will yield in the limit of infinite samples the same value as when averaging over the whole orbit. This seems to be indeed the case, as is shown in figure \ref{diff}, where the difference between the $\lambda=0$ value and the minimal Landau gauge value is shown. As seen, the difference is small, though systematic. However, the deviation is of the same size as the statistical error of the data, and thus this may be a finite statistics effect.

This permits a quite interesting speculation: If a formulation like \pref{quant:bgauge2} is correct, then the case $\lambda=0$ implies minimal Landau gauge. Since $\lambda$ takes the form of an additional gauge parameter, minimal Landau gauge is a fixed point under multiplicative renormalization, and thus the natural result under an unconstrained renormalization evolution. Thus, minimal Landau gauge is the one naturally obtained in the infinite volume and continuum limit. It requires a particular action to stay away from this fixed point.

In functional equations, this is not trivially recovered. Here, formulating the functional equations at finite $\lambda$ and then taking the limit of $\lambda$ to zero would be adequate. Given that the additional term in the path integral \pref{quant:bgauge2} implements a non-local constraint on the ghost propagator equation only, this requires a constraint in this equation. This constraint will force $G(0)$ to the value in the minimal Landau gauge for $\lambda$ going to zero.

\begin{figure}
\includegraphics[width=0.5\textwidth]{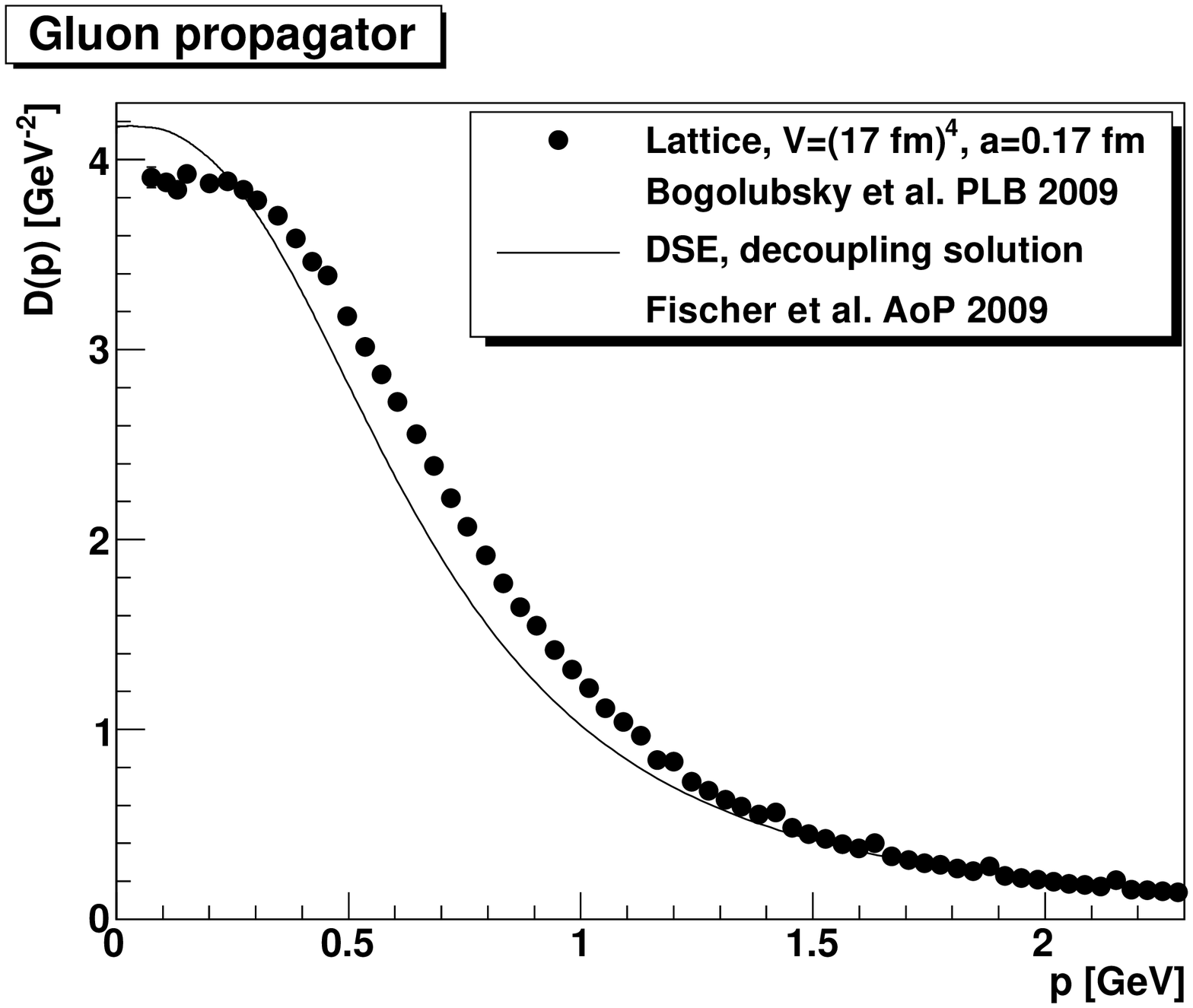}\includegraphics[width=0.5\textwidth]{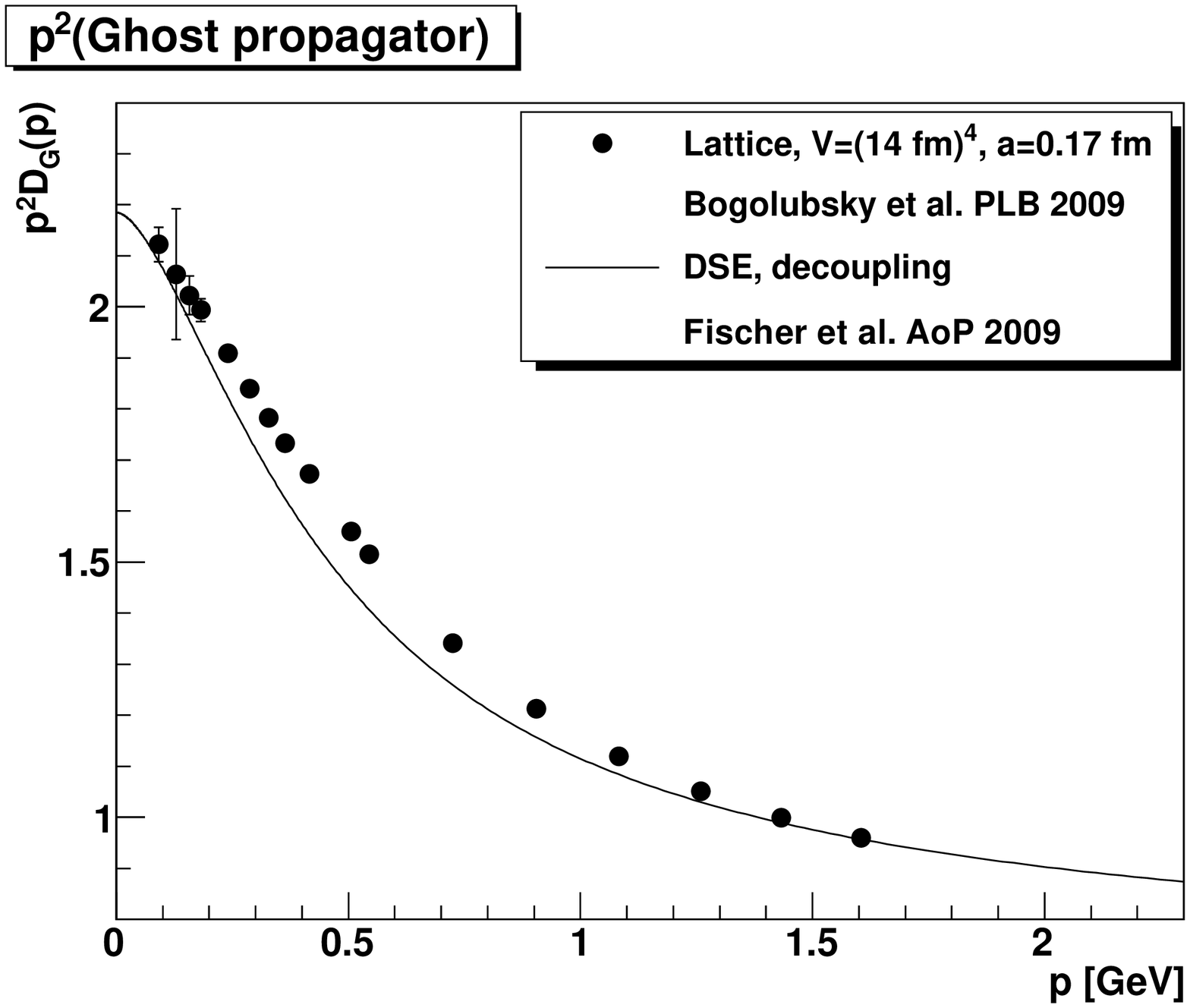}
\caption{\label{prop}Gluon propagator (left panel) and ghost propagator (right panel) from DSEs \cite{Fischer:2008uz} and lattice data \cite{Bogolubsky:2009dc}.}
\end{figure}

Alternatively, instead of taking the limit, it may be possible to just select among the solutions of the unconstrained functional equations by selecting the solution which satisfies the gauge condition $G(0)=B$. For other values of $\lambda$, and thus $B$ and $G(0)$, this procedure should, in principle, hold as well. This prescription has already been used \cite{Fischer:2008uz}, and this yields a rather accurate reproduction of the lattice results for the propagators in minimal Landau gauge using functional methods, as shown in figure \ref{prop}.

What is then about the family of solutions obtained in the functional equations \cite{Maas:2011se,Binosi:2009qm,Boucaud:2011ug,Boucaud:2008ji,Fischer:2008uz,Pawlowski:2003hq}? Since the DSEs in these cases have been derived without an explicit constraint, and are, as argued beforehand, insensitive to the Gribov region without further input, there are a number of scenarios, as discussed in \cite{Maas:2010wb,Maas:2011se}. Assume that the existence of the family, in particular the extremal scaling case \cite{Fischer:2008uz,Alkofer:2000wg}, are not mere artifacts. If this is the case, all of the solutions must be reproducible with other methods. However, since the family of solutions was only found using implicit information, without an explicit implementation of additional constraints in the the functional equations, these solutions may originate from any Gribov region and/or weighting over residual gauge orbits, in the spirit of the discussion in the previous paragraph. Thus, the existence of the family of solutions can only be firmly checked or excluded using lattice gauge theory, if all of these possibilities are checked. For this, it would be helpful to explicitly find the constraints used in the functional equations, in order to be able to reproduce the same calculation on the lattice. In particular, it would be very important to understand whether the scaling case is indeed obtained only when averaging over all Gribov copies in all Gribov regions \cite{vonSmekal:2007ns,vonSmekal:2008es,vonSmekal:2008ws}.

However, if one is only interested in using the propagators to finally obtain gauge-invariant quantities, this can very well be done with any of the family members, in particular the minimal Landau gauge one. If the speculations here are correct, then there is a way to implement the minimal Landau gauge in functional calculations, and this can be done. This ends the speculative part.

In case the range of possible $b$ values shrinks to a point in the continuum and infinite-volume limit, the averaging procedure returns to a flat average, i.\ e., to the case of $\lambda=0$, and all of this is only a rather elaborate framework to recover the flat average over the residual gauge orbit. However, the procedure still remains correct, and all operations performed are permissible.

\section{Where to go from here}

Gauge averaging over Gribov copies as an operative definition in lattice gauge theory is obviously possible. Of course, provided the normalization is correct, these are equally possible gauge choices, and there is no preference from a physical point of view. In fact, at least for those quantities investigated here, the conventional minimal Landau gauge turns out to be equivalent to a particular averaging gauge.

The next step is to investigate for further quantities, foremost correlation functions, what the impact of averaging is. In particular, whether the equivalence to minimal Landau gauge at the fixed point $\lambda=0$ persists. This would establish the concept further concerning lattice gauge theory. Of course, because averaging over all Gribov copies with a certain weight, irrespective of whether the Gribov copies actually receive a different weight or not, precisely defines a treatment of Gribov copies, this is a full resolution of the Gribov-Singer ambiguity.

Concerning the implementation in functional methods, the situation is not yet as simple. Whether the formulation \pref{quant:bgauge2} is indeed correct requires more formal arguments, yet to be developed, if possible at all. The fact that the procedure yields rather good agreement in actual calculations, see figure \ref{prop}, as well as for physical observables \cite{Fischer:2006ub,Fischer:2009gk,Blank:2010pa,Braun:2007bx}, could still be accidental due to the truncations performed. Nonetheless, it offers at least a formulation which, in light of the lattice results, could be a starting point for a more formal solution of the problem. Since a solution is mandatory to formally permit the comparison of continuum and lattice results, a resolution of this problem appears worthwhile to find.

\bibliographystyle{bibstyle}
\bibliography{bib}

\end{document}